\documentclass[aps,pra,showpacs,groupedaddress,superscriptaddress,twocolumn]{revtex4}
\usepackage{amsfonts}
\usepackage{graphicx}
\usepackage{latexsym}
\usepackage{makeidx} 
\usepackage{amsmath} 
\usepackage{amssymb}
\usepackage{graphicx}
\usepackage{color}
\begin{document}

\title{Impact of the dark path on quantum dot single photon emitters in small cavities }
\author{Kenji Kamide}
\email{kamide@iis.u-tokyo.ac.jp}
\affiliation{%
 Institute for Nano Quantum Information Electronics (NanoQuine), University of Tokyo, Tokyo 153-8505, Japan
}
\author{Satoshi Iwamoto}
\author{Yasuhiko Arakawa}%
%\homepage{http://wwwacty.phys.sci.osaka-u.ac.jp/~ogawa/cover.html}
\affiliation{%
 Institute for Nano Quantum Information Electronics (NanoQuine), University of Tokyo, Tokyo 153-8505, Japan
}
\affiliation{%
 Institute of Industrial Science, University of Tokyo, Tokyo 153-8505, Japan
}%
%\affiliation{
\date{\today}% It is always \today, today,
\begin{abstract}
Incoherent pumping in quantum dots (QDs) can create a biexciton state through two paths: via the formation of bright or dark exciton states. 
The latter, dark-pumping, path is shown to enhance the probability of two-photon simultaneous emission, and hence increase $g^{(2)}(0)$ by a factor $\propto 1/\gamma_S$, due to the slow spin relaxation rate $\gamma_S$ in QDs. 
The existence of the dark path is shown to impose a limitation on the single photon (SP) emission process, especially in nanocavities which exhibit a large exciton-cavity coupling and a Purcell enhancement for fast quantum telecommunications. 
\end{abstract} 
\maketitle

{\it Introduction---}
A high quality single photon (SP) source is essential for the realization of secure telecommunications based on the principles of quantum mechanics, such as quantum key distribution (QKD)~\cite{BB84}. 
Semiconductor quantum dots (QDs) are promising candidates for solid-state single photon emitters, because of their well-defined atom-like quantized states ~\cite{Waks, Santori, Takemoto, Kako, Mark}, and a high controllability of their emission wavelength~\cite{Nakaoka}, by means of applied electric and magnetic fields. It is also possible to populate the states of QDs via both current injection and optical excitation. Interactions between QD excitons and photons are also controllable by embedding QDs in optical microcavities such as photonic crystal (Ph-C) nanocavities~\cite{Yablonovitch}, in which cavity quantum electrodynamics (cavity QED) effects have been observed~\cite{Yoshie, Nomura}.  
 
The quality of a SP emitter is quantified by measuring the conditional probability to observe photons at a delay time $\tau$ after a photon counting event, $g^{(2)}(\tau)$~\cite{HBT, Carmichael}. The value of $g^{(2)}(\tau)$ at zero time delay, $g^{(2)}(0)$, should be or be as close as possible to zero to obtain good SP emitters. This is equivalent to minimizing the probability of finding multiple-photon simultaneous emissions.
For application in QKD, a high emission rate is also desired, which can be attained if QDs are combined with optical microcavities~\cite{Englund, Strauf, Ellis, Notomi, Buckley}. 
 
In this brief report, we investigate incoherent pumping processes in a QD SP emitter to find a ``dark path''--- a pumping path from ground to biexciton states via a dark exciton state that can strongly increase $g^{(2)}(0)$, thus imposing a limit on the available single photon purity, especially in small cavities with a small mode volume, $V_{\rm mode}$, and a large exciton-cavity coupling, $g \propto 1/\sqrt{V_{\rm mode}}$. 
Several ways to reduce the effect of the dark path are also mentioned. 
In the following analysis, we define $\hbar = 1$ for simplicity.

%An enhancement of the two-photon emission probability in QD systems is in a striking contrast to an ideal atomic two-level system, that arises in exchange for the high controllability and high coupling efficiency to photons.

\begin{figure}[tb]
\begin{center}  
\includegraphics[scale=0.26]{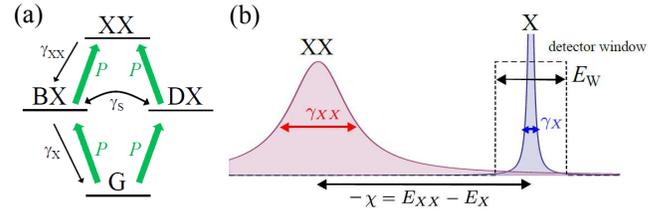} {}
\end{center}
\vspace{-5mm}
\caption{\label{fig1} (a) Ground, exciton, and biexciton states (G, BX, DX, and XX), pumping ($P$) and relaxation processes ($\gamma_X, \gamma_{XX}$, and $\gamma_S$) in an undoped QD described in the text. (b) Emission spectral profiles of the biexciton and exciton recombination transitions. The spectral filter used for the photodetection (with a detection window $E_W>\gamma_X$) is also shown. }
\end{figure}

{\it Impact of the dark path in QD SP emitters without cavities --- }
In order to see how the dark path increases the multiple photon emission probability, we first study a simple phenomenological model for a QD SP emitter without cavity coupling.
Here, an undoped QD is pumped incoherently and continuously under a charge-neutral condition. 
The QD states relevant to our study are restricted to the neutral states with up to two electron-hole pairs as shown in Fig.~\ref{fig1} (a).
Excited carriers injected at high energy levels become trapped in the lowest QD level after fast relaxation from a continuum above the band gap and excited trapped states (which are truncated in our model).
The whole process of injection and relaxation of the carriers (electron-hole pairs) to the lowest QD level is described by a single pumping rate $P$.
Depending on the spins of carriers, the bright and dark exciton states (BX/DX) are randomly generated from the initial empty state (G) with the same rate $P$.  
Successive creations of two electron-hole pairs further excite the system to the biexciton state, XX. The single photon emission process is mainly governed by the recombination transition from BX to G with a spontaneous emission rate $\gamma_X$, and the two photon emission process by the cascaded emission from XX to G via BX (the energy difference between biexciton and single-exciton emissions is the binding energy $-\chi=E_{XX}-E_{X}$). 
In QDs, spin relaxation processes between BX and DX (the rate $\gamma_S$) are usually slow and often neglected
compared to the other dynamics~\cite{Paillard, Smith}, however they must be considered carefully to evaluate $g^{(2)}(0)$ as shown below.  

The rate equations for the populations at each QD levels are $\dot{\rho}_G=-2P \rho_G +\gamma_X \rho_{BX}$, 
$\dot{\rho}_{DX}=-(\gamma_S +P) \rho_{DX} +P \rho_{G} +\gamma_S \rho_{BX}$, $\dot{\rho}_{BX}=-(\gamma_X +\gamma_S +P) \rho_{BX} +P \rho_{G} +\gamma_{XX} \rho_{XX}+\gamma_S \rho_{DX}$, and $\dot{\rho}_{XX}=-\gamma_{XX} \rho_{XX} + P \rho_{DX}+ P \rho_{BX}$, 
for which the steady state satisfies 
\begin{eqnarray}
\rho_{XX} &=&\left(1+\frac{\gamma_X}{4\gamma_S} \right)\frac{4P^2}{\gamma_{X}\gamma_{XX}}+\mathcal{O} (P^3), \label{eq:rhoXX}\\
\rho_{BX}& =&2P/\gamma_X+\mathcal{O} (P^2), \label{eq:rhoBX}\\
\rho_{DX}&=&(\gamma_S^{-1}+2\gamma_X^{-1})P +\mathcal{O} (P^2), \label{eq:rhoDX}
\end{eqnarray}
 in the linear regime at weak pumping. From this result, neglecting a factor of order unity related to the emission profiles, the normalized second order correlation function is given by
$g^{(2)}(0)=f_{XX}\times (\gamma_{X} \gamma_{XX} \rho_{XX})/  ( \gamma_X \rho_{BX})^2 $~\cite{note1}. 
Here, 
\begin{eqnarray}
f_{XX} &\equiv& \frac{1}{\pi} \int_{E_X-E_W/2}^{E_X+E_W/2} \frac{ \gamma_{XX}}{\gamma_{XX}^2+(\omega-E_{XX})^2} {\rm d} \omega 
\end{eqnarray}
is the probability of finding a photon emitted from $XX$  within the spectral range of the detector window  $E_X-E_W/2<\omega< E_X+E_W/2$ as shown in Fig.~\ref{fig1}(b). Upon making the assumptions that $\chi \gg E_W$, $\chi \gg \gamma_{XX}$ and $E_W> \gamma_X$, $f_{XX}$ can be expressed as
\begin{eqnarray}
f_{XX} & \sim& \pi^{-1} (\gamma_{XX}/\chi)^2 (E_W/\gamma_{XX}),
\end{eqnarray}
  and hence, in the linear regime
\begin{eqnarray}
g^{(2)}(0)=\frac{1}{\pi} \left(\frac{\gamma_{XX}}{\chi} \right) \left(\frac{ E_{W}}{\chi} \right) \left(1 +\frac{\gamma_X}{4 \gamma_S} \right). \label{eq:g2simple}
\end{eqnarray}

The expression for $g^{(2)}(0)$ contains a factor $\left(1+\frac{\gamma_{X}}{4\gamma_S}\right)$ that becomes large when the spin relaxation is very slow $\gamma_S \ll \gamma_X$~\cite{note2}. 
The enhancement of $g^{(2)}(0)$ occurs due to an unwanted excitation of state XX through a ``dark path''---a path via the excitation of DX. 
This can can be verified by testing using another model without the dark path: We find that $g^{(2)}(0)=\pi^{-1}\gamma_{XX} E_{W} \chi^{-2} $ if the DX state and the ``dark'' pumping path are not present (this result is also obtained when $\gamma_S \gg \gamma_X$ in Eq.~(\ref{eq:g2simple}).) 
The above scenario can also be verified by considering the steady state population in Eq.~(\ref{eq:rhoBX}) and Eq.~(\ref{eq:rhoDX}): the dark state population $\rho_{DX}$ and the ratio $\rho_{DX}/\rho_{BX}$ diverge as $\gamma_S/\gamma_X \to 0$ so that the production rate of $\rho_{XX}$ is largely enhanced when the dark pumping path is present. 

Following the simple discussion above, we find that the dark path can act as a bottleneck when trying to purify the single photon generation in QDs, since the spin relaxation process is usually slower (typically 1-10 ns timescale) than the other processes~\cite{Paillard,Smith}: radiative recombination and dephasing, in QD systems. 

The strong impact of a dark pumping path on $g^{(2)}(0)$ has been shown for many two-level atoms or many QDs by Temnov {\it et al.}~\cite{Temnov}, where the connection with  cooperative spontaneous emission and superradiance are also discussed. Even though the system considered here is a single emitter, similar physics does exist and can reduce the quality of a SP emitter.

The interpretation of the other parts in Eq.~(\ref{eq:g2simple}), $\frac{1}{\pi} \left(\frac{\gamma_{XX}}{\chi} \right) \left(\frac{ E_{W}}{\chi} \right) $, is physically clear; $g^{(2)}(0)$ can be made small if the biexciton and exciton emissions are energetically separated (large $\chi$)~\cite{Langbein,Trotta} and the detector window $E_W$ is small, by which the probability of counting unwanted photons fed by XX state into the window can be reduced.

{\it Impact of the dark path on high-speed SP emitters with cavities---}
We next shift our discussion to how the use an optical microcavity~\cite{Englund, Strauf, Notomi} (which increases the SP emission rate by Purcell effect) affects the single photon purity. 
The realization of high-quality SP sources in microcavities is an important issue, and a number of experiments have been reported on the fabrication and evaluation of such devices (for a recent review, see \cite{Buckley}). 
Hence, it is important to understand what affect such a dark path may have on QDs in microcavities.
From the result obtained in the previous section, (Eq.~(\ref{eq:g2simple})), one may guess that the Purcell enhancement in $\gamma_{X}$ would result in the increase in $g^{(2)}(0)$, and hence that the ``dark path'' pumping might be more troublesome than in a system without a cavity. However, the discussion without cavity cannot be applied directly to this case. 
Here we investigate the effect of a microcavity within the cavity QED framework with quantum master equations (QME)~\cite{Carmichael}, treating the dynamics of cavity mode photons as well. 

We consider a system that consists of carriers inside a QD and photons interacting inside a cavity. 
Six electronic configurations of the QD are considered~\cite{note3}: 
an empty state, $| G \rangle $, two bright exciton states, $| BX1 \rangle =e_\uparrow^\dagger h_\downarrow^\dagger | G \rangle $ and $| BX2 \rangle =e_\downarrow^\dagger h_\uparrow^\dagger | G \rangle $, two dark exciton states, $| DX1 \rangle =e_\uparrow^\dagger h_\uparrow^\dagger | G \rangle $ and $| DX2 \rangle =e_\downarrow^\dagger h_\downarrow^\dagger | G \rangle $, and a biexciton state, $| XX \rangle =e_\uparrow^\dagger e_\downarrow^\dagger h_\uparrow^\dagger h_\downarrow^\dagger | G \rangle $, where $e_\sigma$ and $h_\sigma$ ($e_\sigma^\dagger $ and $h_\sigma^\dagger $) are annihilation (creation) operators of electrons and holes with spin $\sigma=\uparrow, \downarrow$ in their respective lowest energy levels of the QD.

The number of cavity photons is given by $a^\dagger a$, where $a$  and $a^\dagger$ are the photon annihilation and creation operators, respectively.
Assuming the frequencies of the cavity ($\omega_C$) and exciton ($\omega_X$) are tuned to resonance, $\omega_C=\omega_X \equiv \omega_0$, the Hamiltonian of the coupled QD-cavity system~\cite{Yamaguchi} can be written as
\begin{eqnarray}
H&=&\omega_0 N_{\rm tot} -\chi |XX \rangle \langle XX|  \nonumber \\
&+& \sum_{i=2,3}  g_Xa^\dagger  | G \rangle \langle i| +g_{XX} a^\dagger  | i \rangle \langle XX| +{\rm h.c.},
\end{eqnarray}
where $N_{\rm tot}=\sum_{\sigma=\uparrow,\downarrow} (e_\sigma^\dagger e_\sigma+h_\sigma^\dagger h_\sigma  )/2 +a^\dagger a$ is the total excitation number, and we put $g_X=g_{XX} \equiv g$ for simplicity in the simulations.  
Assuming the dynamics in the environment (pump and decay baths outside the coupled QD-cavity system) are fast and uncorrelated, the time evolution of the system density matrix is given by Markovian QME, $\frac{\rm d}{{\rm d}t}\rho= i[\rho, H]+\mathcal{L} \rho$~\cite{Carmichael} , where
\begin{eqnarray}
\mathcal{L} \rho&=&\left( \kappa\mathcal{L}_{a}+P\sum_{\sigma,\sigma'} \mathcal{L}_{h^\dagger_{\sigma '} e^\dagger_\sigma  } +\sum_{\sigma} \left( \gamma_{sp} \mathcal{L}_{ e_\sigma h_{-\sigma} } + \Gamma_{\rm ph} \mathcal{L}_{e_\sigma^\dagger  e_{\sigma} }  \right. \right. \nonumber \\
&& \left. \left.  + \Gamma_{\rm ph} \mathcal{L}_{h_\sigma^\dagger  h_{\sigma} } + \gamma_{S}^e  \mathcal{L}_{ e_\sigma^\dagger  e_{-\sigma}}+\gamma_{S}^h \mathcal{L}_{ h_\sigma^\dagger  h_{-\sigma} } \right) \right) \rho. \qquad \label{eq:QME}
\end{eqnarray}
Using the standard notation for superoperators, $\mathcal{L}_{A}\rho \equiv \frac{1}{2}(2A \rho A^\dagger -A^\dagger A \rho-\rho A^\dagger A)$, we consider the following situation; the decay of the injected electron-hole pairs is dominated by the spontaneous emission into the cavity mode (whose couplings are $g_{X}$ and $g_{XX}$) and free space (with a rate $\gamma_{sp}$)~\cite{Johansen}; photons decay out of the cavity with rate $\kappa$; the polarizations suffers dephasing with rate $\Gamma_{\rm ph}$; the spin flip of electrons and holes (with the rates $\gamma_S^e$ and $\gamma_S^h$) results in the transitions between dark and bright exciton states with a rate $\gamma_S=\gamma_S^e+\gamma_S^h$.

One could determine the non-equilibrium steady state numerically by a long-time evolution of the system, $\rho_{\infty}$, from an initial vacuum state $\rho _{0}=|G \rangle \langle G| \otimes |0\rangle \langle 0|_{\rm cav}$ after the pumping is switched on $P>0$ at $t=0$. 
The obtained density matrix, $\rho_{\infty}$, can then be used to determine the photon number $\langle a^\dagger a \rangle$ and 
$g^{(2)}(0)= \langle a^\dagger  a^\dagger a  a \rangle / | \langle a^\dagger a \rangle |^2 $. 
However, we choose an alternative analytic approach to find the photon correlation functions, which is allowed in a linear regime at weak pump rate, and can be performed by a perturbation method~\cite{Agarwal, delValle, Gartner}. 
This analytic approach brings clear insights of the physics and greatly reduces the calculation time to obtain the properties of photons as a function of numbers of parameters ($\kappa, P, \gamma_{sp}, \Gamma_{\rm ph}, \gamma_S, \chi$). 
See the supplementary material for details of the calculation~\cite{suppleAB}. 

\begin{widetext} 
\begin{figure*}[tb] 
\begin{center}
\includegraphics[scale=0.5]{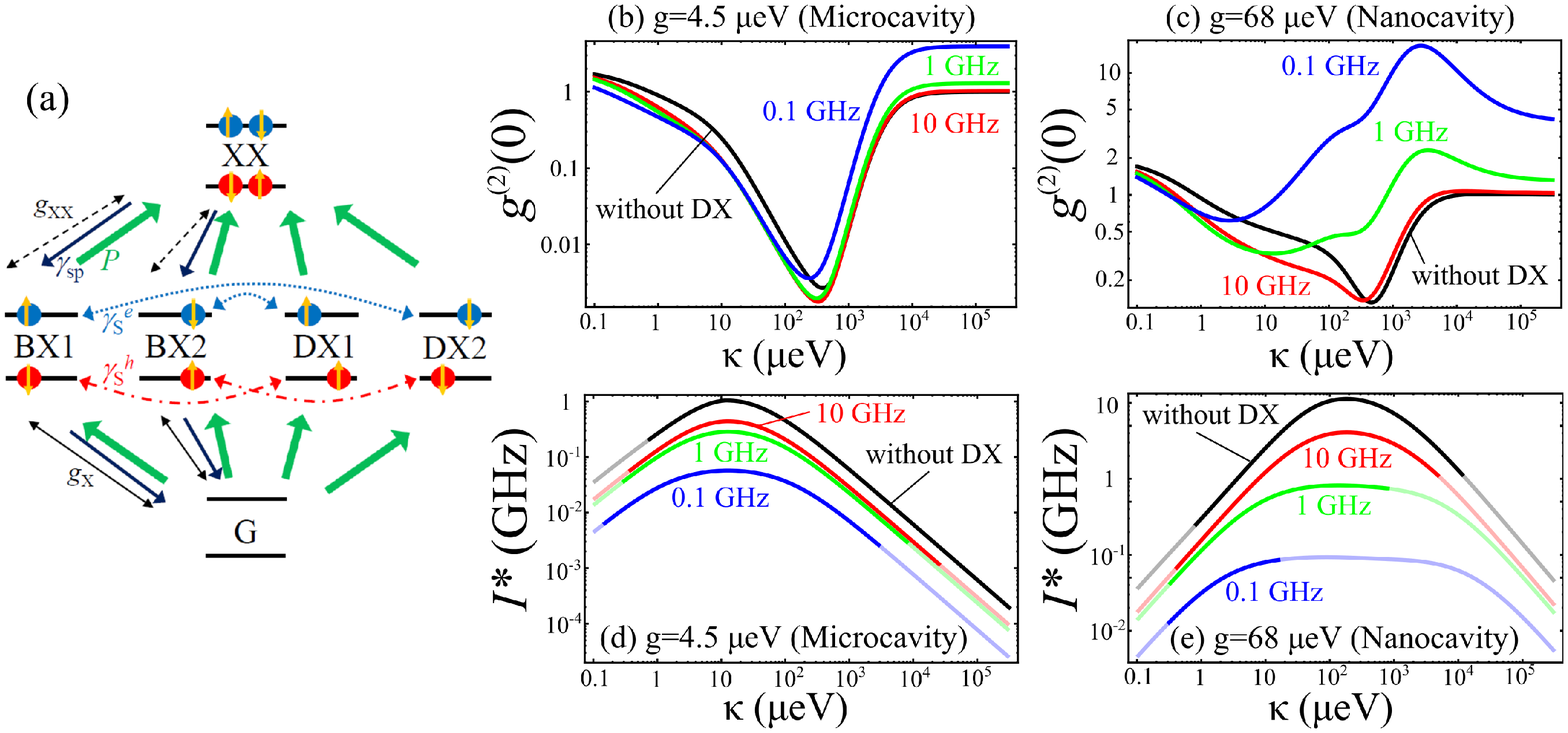} {} 
\vspace{-3 mm}
\caption{\label{fig2}  (a) A QD model with full consideration of carrier spin configurations for SP emitters with cavities. The image shows the Empty, exciton, and biexciton states (G, BX1, BX2, DX1, DX2, and XX), pumping ($P$), spontaneous emission into free space ($\gamma_{sp}$), spin relaxation ($\gamma_S$), exciton-cavity and biexciton-cavity couplings ($g_X, g_{XX}$) in an undoped QD as described in the text. We used $\gamma_S \equiv \gamma_S^e+\gamma_S^h$ and $g_X=g_{XX} \equiv g$ in the simulation. In the QME, Eq.~(\ref{eq:QME}), a dephasing rate $\Gamma_{\rm ph}$ is also considered. 
(b,c) $g^{(2)}(0)$, and (d,e) a crossover photon emission rate, $I^\ast $, plotted as a function of the cavity loss $\kappa$ for (b,d) a microcavity with $V_{\rm mode} \sim \ 15 \times 15 \times 1 \mu {\rm m}^3$ and $g = 4.5 \ \mu {\rm eV}$, and for (c,e) a ultrasmall nanocavity with $V_{\rm mode} \sim \ 1 \times 1 \times 1 \mu {\rm m}^3$ and $g = 68 \ \mu {\rm eV}$~\cite{Strauf, Nomura} . 
Three spin relaxation rates, $\gamma_S=0.1, \ 1, \ 10\ {\rm GHz}$, are used as indicated in the plots.
Typical values of InAs QDs are chosen for the spontaneous emission rate, dephasing rate, and biexciton binding energy: $\gamma_{sp}=0.77 \ \mu {\rm eV}=1/(0.85 \ {\rm ns} )$, $\Gamma_{\rm ph}=15 \ \mu {\rm eV}$ and $\chi = 2$ meV for all plots.  
Pale parts of the plots in (d,e) are corresponding to the regime where $g^{(2)}(0)>1$ in (b,c) and $I^\ast$ cannot be a measure of the maximum available emission rate as a quantum light source~\cite{suppleAB}. }
\end{center}
\end{figure*} 
\end{widetext}

Figures~\ref{fig2} (b,c) and (d,e) show the calculated $g^{(2)}(0)$ and the crossover photon emission rate $I^\ast$ (defined as the photon emission rate $\kappa \langle a^\dagger a \rangle$ at the crossover pump rate $P$ between the linear and nonlinear regime~\cite{suppleAB}) as a function of $\kappa$ for a microcavity and a nanocavity, respectively.  $I^\ast$ indicates the maximum available photon emission rate with $g^{(2)}(0)$ being kept low when $g^{(2)}(0)$ is less than unity in the weak pump limit.  
Figures~\ref{fig2} (b,d) are the calculation results for a microcavity structure as in \cite{Strauf} with $V_{\rm mode} \sim 15 \times 15 \times 1 \mu {\rm m}^3$ and $g = 4.5 \ \mu {\rm eV}$. Figures~\ref{fig2} (c,e) show the results for an ultrasmall nanocavity structure, e.g. a Ph-C nanocavity as in \cite{Nomura,note4}, with $V_{\rm mode} \sim \ 1 \times 1 \times 1 \mu {\rm m}^3$ and $g = 68 \ \mu {\rm eV}$.
In each plot, three spin relaxation rates, $\gamma_S=0.1, \ 1$, and $10$ GHz, are examined. 
The spontaneous emission rate $\gamma_{sp}=0.77 \ \mu {\rm eV}=1/(0.85 \ {\rm ns} )$, dephasing rate $\Gamma_{\rm ph}=15 \ \mu {\rm eV}$, and the biexciton binding energy $\chi = 2$ meV used in all calculations are typical values for InAs QDs.

Interestingly, the figures exhibit an optimal $\kappa \equiv  \kappa_{\rm opt, 1}$ minimizing $g^{(2)}(0)$ ($\kappa_{\rm opt, 1} \sim 300 \ \mu$eV for both cavities with $\gamma_S=10\ {\rm GHz}$.) 
They also exhibit an optimal $\kappa$ ($\equiv \kappa_{\rm opt, 2}$), which maximizes $I^\ast$. 
This is in a remarkable contrast to cavity QED with a two-level atom~\cite{delValle, Gartner}, in which $g^{(2)}(0)$ decreases from two down to zero as $\kappa$ increases from the good-cavitiy to bad-cavity limits.
Thus, we understand that the existence of optimal cavity losses arises due to the existence of the biexciton (multiple exciton) state. 
The optimal loss is found in a range $g < \kappa_{\rm opt,1} < \chi$. This is explained by connecting the two limits: (i) For good cavities with $\kappa \ll g$, photons accumulate in the cavity resulting in the non-negligible multiple-photon probability resulting in a decreasing $g^{(2)}(0)$ with increasing $\kappa$. (ii) In the weak coupling regime $\kappa \gg  g$, the rate of exciton and biexciton transitions into the cavity mode ($W_X=2 g^2/\kappa$, $W_{XX}=2 g^2 \kappa/(\kappa^2 + \chi^2)$ become of similar order for $\kappa \gg \chi$, leading to increased cascaded two-photon emission, and an increasing $g^{(2)}(0)$ with $\kappa$ (See Sec.~\ref{suppleC} of the supplements~\cite{suppleC}.)

\begin{figure}[tb] 
\begin{center}  
\includegraphics[scale=0.53]{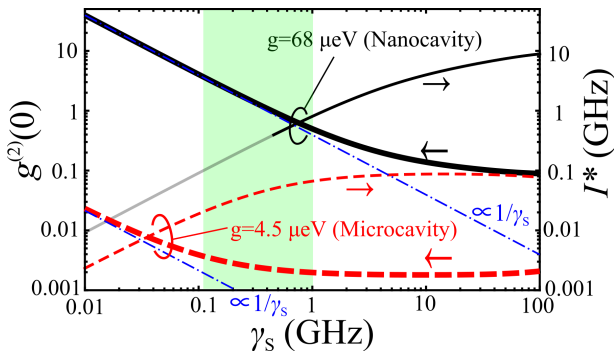} {}
\vspace{-1mm}
\caption{\label{fig3} Second-order coherence $g^{(2)}(0)$ (thick) and a crossover photon emission rate $I^\ast$ (thin) as a function of spin relaxation rate $\gamma_S$ for a microcavity (red dashed) and a nanocavity (black solid) with $\kappa=300 \ \mu {\rm eV}$. Other parameters for the microcavity and nanocavity are the same as in Fig.~\ref{fig2}. 
Dot-dashed lines: guides for the eye on $g^{(2)}(0) \propto 1/\gamma_S$. The pale part of the $I^\ast$ plot for the nanocavity corresponds to the region where $g^{(2)}(0)>1$. The green shaded area indicates a typical parameter range for InAs QDs ($0.1 \ {\rm GHz}<\gamma_S<1 \ {\rm GHz}$).}
\end{center}
\end{figure} 

An Important finding is that for the larger microcavity in Fig.~\ref{fig2} (b, d), The values of $g^{(2)}(0)$ at the minima do not change so much as the spin relaxation rate changes ($\gamma_S=0.1-10$ GHz). On the other hand, for the smaller nanocavity in Fig.~\ref{fig2} (c, e), one finds a large enhancement in $g^{(2)}(0)$ for the slow spin relaxation rate ($\gamma_S < 1$ GHz). 
QDs in a nanocavity with $\gamma_S < 1$ GHz cannot be considered as a SP emitters since $g^{(2)}(0)$ never falls below 0.5 (unless some means were taken as mentioned below)! 
Therefore, the impact of the ``dark path'' on the SP emission is stronger in smaller cavities, which agrees with the simple intuitive guess made at the beginning of this section.
For comparison, we also show calculated results for the QD model without the dark states and dark paths~\cite{suppleC} (solid black curves in Fig.~\ref{fig2} (b)-(e).) 
An increase in $g^{(2)}(0)$ at $\kappa = \kappa_{\rm opt,1}$ as  $\gamma_S$ changes from  $1$ to $0.1$ GHz is clear in Fig.~\ref{fig2} (c).
A reduction of $I^\ast$ due to the dark path, claimed by Strauf {\it et al.}~\cite{Strauf}, is also found in Fig.~\ref{fig2} (d,e), and the reduction is much more pronounced for the nanocavity in Fig.~\ref{fig2}~(e) (The nano cavity exhibits a more than 10-fold decrease at $\kappa = \kappa_{\rm opt,2}$ for $\gamma_S=1$ GHz.)

Figure \ref{fig3} shows the $\gamma_S$-dependence of $g^{(2)}(0)$ and $I^\ast$ for a cavity loss $\kappa=300 \ \mu {\rm eV}$ (chosen around $\kappa=\kappa_{\rm opt, 1}$). 
For both cavities,  $g^{(2)}(0)$ is proportional to $1/\gamma_S$ at small $\gamma_S$ due to the enhanced dark path effect. 
The $1/\gamma_S$-dependency and an observation that the dark pumping path reduces the quality of QD SP emissions for slow spin relaxations originate from the same physics explained above.
However, for the case of the microcavity, the $1/\gamma_S$-dependency is not found in the typical parameter regime ($0.1 \ {\rm GHz}<\gamma_S<1 \ {\rm GHz}$: the shaded region in Fig.~\ref{fig3}), meaning the dark path rarely causes the unwanted effect of multiple photon emission.
On the other hand, this effect is always important in the small Ph-C nanocavity, since the $1/\gamma_S$-dependency is found in the typical regime.

{\it Summary---} The strong impact of the ``dark path''---
a pumping path via dark exciton states to biexciton states---on the quality of SP emitters has been shown to exist in QD systems. 
The observed increase in the value of $g^{(2)}(0)$ indicates the dark path can reduce quality of QD SP emitters,  especially those situated in small cavities like nanocavities. 
The effects of the dark path were investigated under cw excitation conditions, but will also manifest themselves under long-pulse excitation. 
The effect of short-pulse excitation will be discussed elsewhere.     

Finally, we mention several ways to reduce the impact of the dark path for an application purpose. 
(i) The use of a charged exciton state $X_\pm$ where there is no dark path~\cite{Strauf,note4}. 
(ii) The use of resonant and coherent laser excitation~\cite{Paillard2, Muller, Vamivakas, Ates2, Englund2} which automatically selects to create bright states not the dark states. (iii) Enhancement of the spin relaxation rate~\cite{Smith} to suppress the unwanted multiple photon emission. (iv) Selection of QDs with a large biexciton binding energy~\cite{Langbein} in order to limit the spectral overlap of cascaded photons.    
(v) The use of a spectral filter to further reduce the unwanted output from XX emission can partly reduce the impact of the dark path if the filter bandwidth were optimally selected~\cite{Kamide}.

{\it Acknowledgements---}
We thank Y. Ota, M. Holmes, and S. Kako for useful comments and discussions.
This work is supported by the Project for Developing Innovation Systems of MEXT.

\clearpage
\newpage 
%%%%%%
%%%%%%
%%%%%%
%%%%%%
%%%%%%
%%%%%%
\section{\label{supple} Supplementary information }
\subsection{\label{suppleA} Derivation of coupled equations of motion for correlation functions}
The derivation of the the closed set of coupled equations of motion for correlation functions for two-level laser systems is given  in Ref.~\cite{Agarwal, delValle, Gartner}. For our QD model (shown in Fig.~\ref{fig2}) a more complicated but similar analysis can be performed. 
To simplify the discussion, we assume that $g_X=g_{XX} \equiv g$, which does not affect the underlying physics 
(In real systems, $g_X \ne g_{XX}$, whose effect could be effectively incorporated by introducing an additional shift in $\chi=E_{X}-E_{XX}$.)
The correlation functions that we need to consider are 
\begin{eqnarray}
{\bf G}_k &\equiv& {\rm Tr} \left(  | G \rangle \langle G | (a^\dagger)^k a^k  \rho  \right), \\
{\bf BX}_k &\equiv& {\rm Tr} \left(  |BX1 \rangle \langle BX1 | (a^\dagger)^k a^k  \rho  \right) \nonumber \\
&=&{\rm Tr} \left(  |BX2 \rangle \langle BX2 | (a^\dagger)^k a^k  \rho  \right),  \label{eq:BXk} \\
{\bf DX}_k &\equiv& {\rm Tr} \left(  |DX1 \rangle \langle DX1 | (a^\dagger)^k a^k  \rho  \right) \nonumber \\
&=&{\rm Tr} \left(  |DX2 \rangle \langle DX2 | (a^\dagger)^k a^k  \rho  \right), \label{eq:DXk} \\
{\bf XX}_k &\equiv& {\rm Tr} \left(  | XX \rangle \langle XX | (a^\dagger)^k a^k  \rho  \right),  \\
{\bf t}_{X,k+1} &\equiv& -ig \  {\rm Tr} \left(  |G \rangle \langle BX1 | (a^\dagger)^{k+1} a^{k}  \rho  \right) \nonumber \\
&=&-ig \  {\rm Tr} \left(  |G \rangle \langle BX2 | (a^\dagger)^{k+1} a^{k}  \rho  \right), \label{eq:tXk} \\
{\bf t}_{XX,k+1} &\equiv& -ig \  {\rm Tr} \left(  |BX1 \rangle \langle XX | (a^\dagger)^{k+1} a^{k}  \rho  \right) \nonumber \\
&=&-ig \ {\rm Tr} \left( |BX2 \rangle \langle XX | (a^\dagger)^{k+1} a^{k}
 \rho  \right), \quad  \label{eq:tXXk} \\
{\bf R}_{k+2} &\equiv& g^2 \  {\rm Tr} \left(  | G \rangle \langle XX | (a^\dagger)^{k+2} a^{k}  \rho  \right), \\
{\bf BX12}_k &\equiv&  {\rm Tr} \left(  | BX1 \rangle \langle BX2 | (a^\dagger)^k a^{k}  \rho  \right) \nonumber \\
&=& {\rm Tr} \left(  | BX2 \rangle \langle BX1 | (a^\dagger)^k a^{k}  \rho  \right), \label{eq:BX12k}
\end{eqnarray}
for $k =0,1,\cdots, \infty$. Equations~(\ref{eq:BXk}), (\ref{eq:DXk}), (\ref{eq:tXk}), (\ref{eq:tXXk}), and (\ref{eq:BX12k}), are reformed based the spin inversion symmetry in the system.
For the $k$-th photon correlation function, the following relationship is fulfilled:
\begin{eqnarray}
{\bf N}_k &\equiv&   {\rm Tr} \left( (a^\dagger )^k a^k  \rho  \right)  \nonumber \\
&=& {\bf XX}_k +2{\bf BX}_k+2{\bf DX}_k+{\bf G}_k.
\end{eqnarray}
The relationship is used to obtain the second order coherence at zero time delay, $g^{(2)}(0)={\bf N}_2/{\bf N}_1^2$. 
Derivation of the equation of motion for the correlation function is straightforward using $\frac{\rm d}{{\rm d}t}\langle O \rangle = {\rm Tr} \left( O \frac{\rm d}{{\rm d}t} \rho \right)$ with the QME, $\frac{\rm d}{{\rm d}t}\rho= i[\rho, H]+\mathcal{L} \rho$. Their explicit form is as follows:
\begin{widetext}
\begin{eqnarray}
\frac{d}{dt} {\bf G}_k &=&-(\kappa k +4P ){\bf G}_k +2 \gamma_{sp}{\bf BX}_{k}+4 \ {\rm Re}\left({\bf t}_{X,k+1}\right)+4k \ {\rm Re}\left( {\bf t}_{X,k}\right), \label{eq:EOM1} \\
\frac{d}{dt} {\bf DX}_k&=&-(\kappa k +P+ \gamma_S){\bf DX}_k + \gamma_S {\bf BX}_k+P {\bf G}_k, \\
\frac{d}{dt} {\bf BX}_k&=&-(\kappa k +P+ \gamma_{sp}+\gamma_S){\bf BX}_k +P \ {\bf G}_k+\gamma_{sp} {\bf XX}_k+\gamma_S {\bf DX}_k \\
&&-2 \  {\rm Re} \left({\bf t}_{X,k+1} \right)+2 \  {\rm Re} \left({\bf t}_{XX,k+1} \right)+2k \  {\rm Re} \left({\bf t}_{XX,k} \right), \\
\frac{d}{dt} {\bf XX}_k 
&=& -(\kappa k +2  \gamma_{sp}){\bf XX}_k +2P \ {\bf BX}_k +2P \ {\bf DX}_k 
-4 \  {\rm Re} \left({\bf t}_{XX,k+1} \right), \\
\frac{d}{dt} {\bf t}_{X,k}&=& -\xi_k \ {\bf t}_{X,k} +g^2 \ {\bf BX}_{k}+g^2 k \ {\bf BX}_{k-1}-g^2 \  {\bf G}_{k} \\
&& -{\bf R}_{k+1}+(k-1) {\bf R}_{k} +g^2 \ {\bf BX12}_{k}+g^2 k \ {\bf BX12}_{k-1}, \\
\frac{d}{dt} {\bf t}_{XX,k}&=& (i \chi-\eta_k) \  {\bf t}_{XX,k} +g^2 \  {\bf XX}_{k}+g^2 k \  {\bf XX}_{k-1} -g^2 {\bf BX}_{k} +{\bf R}_{k+1}-g^2 {\bf BX12}_{k}, \\
\frac{d}{dt}{\bf R}_{k}&=& \left( i \chi -\zeta_k \right){\bf R}_{k}+2g^2 \ {\bf t}_{X,k}-2g^2 \ {\bf t}_{XX,k}-2g^2 k \ {\bf t}_{XX,k-1}, \\
\frac{d}{dt} {\bf BX12}_k &=& -\theta_k \ {\bf BX12}_k -2 \ {\rm Re} \left( {\bf t}_{X,k+1}  \right) +2 \ {\rm Re} \left( {\bf t}_{XX,k+1}  \right) +2k \ {\rm Re} \left( {\bf t}_{XX,k}  \right), \label{eq:EOM8}  
\end{eqnarray}
\end{widetext}
where we define the exciton spin relaxation rate $\gamma_S \equiv \gamma_S^{e}+\gamma_S^{h}$ and the decay coefficients are given by 
\begin{eqnarray}
\xi_k & \equiv & \kappa \left( k-\frac{1}{2} \right) +\frac{5}{2}P+\frac{\gamma_{sp}}{2}+\Gamma_{\rm ph}+\frac{\gamma_S}{2}, \\
\eta_{k} & \equiv & \left( k -\frac{1}{2} \right) \kappa +\frac{P}{2}+\frac{3}{2}\gamma_{sp} +\Gamma_{\rm ph}+\frac{\gamma_{S}}{2} , \\
\zeta_k & \equiv & \left( k-1 \right)\kappa +2 P +\gamma_{sp}+2 \Gamma_{\rm ph}, \\
\theta_{k} & \equiv & \kappa k +P+\gamma_{sp} +2 \Gamma_{\rm ph}+\gamma_{S} .
\end{eqnarray}
In the steady state ($\frac{\rm d}{{\rm}t} \langle \cdot \rangle =0$), a balance relation is obtained for $k \ge 1$,
\begin{eqnarray}
\kappa \ {\bf N}_k = 4 \ {\rm Re} \left( {\bf t}_{X,k} \right)+4  \ {\rm Re} \left( {\bf t}_{XX,k} \right). \label{eq:balance}
\end{eqnarray}   
For $k=1$, the physical meaning is clearly the balance equation between ``loss (l.h.s.)'' and ``gain (r.h.s.)''. The first and second terms on the r.h.s. correspond to the optical gain contributions from the exciton (BX1/BX2 $\to$ G) and biexciton (XX $\to$ BX1/BX2) transitions.

\subsection{\label{suppleB} Perturbative analytic solution in the linear regime}
The coupled equations of motion in Eqs.~(\ref{eq:EOM1})-(\ref{eq:EOM8}) produce an infinite series of equations from $k=0$ to $\infty$. These equations can be solved numerically if the high-order correlation functions are ignored by introducing a photon-number cutoff $n_{\rm max}$ i.e. by putting ${\bf O}_k=0$ for $k \ge n_{\rm max}$. However, this is unreasonable for the evaluation of the first ($k=1$) and second order ($k=2$) correlation functions in the linear regime at weak pumping.
Here, we give a brief introduction to the perturbative treatment which we perform to evaluate the second-order coherence. This treatment allows us to calculate analytic expressions and provides more insight into the physics, and the dependence of QD populations, mean photon number, and $g^{(2)}(0)$ on numbers of parameters ($\kappa, P, \gamma_{sp}, \Gamma_{\rm ph}, \gamma_S, \chi$).

By assuming the following pump rate dependence, (for the linear regime):
\begin{eqnarray*}
&& {\bf G}_k=\mathcal{O}(P^{k}), \ {\bf BX}_k, {\bf DX}_k=\mathcal{O}(P^{k+1}), \\
&&{\bf XX}_k=\mathcal{O}(P^{k+2}),  \ {\bf t}_{X,k}=\mathcal{O}(P^{k}), \ {\bf t}_{XX,k}=\mathcal{O}(P^{k+1}), \\
&& {\bf R}_k=\mathcal{O}(P^{k}), \ {\bf BX12}_k=\mathcal{O}(P^{k+1}),
\end{eqnarray*}
we are able to obtain these correlation functions perturbatively in powers of small $P$.  
Hereafter, the perturbation series of the $k$-th order correlation function ${\bf O}_k$ is defined as
\begin{eqnarray*}
{\bf O}_k=\sum_{j} {\bf O}_k^{(j)},
\end{eqnarray*} 
where the $j$-th component is proportional to $P^j$.
In the zeroth order, ${\bf G}_0^{(0)}=1$, and all other quantities are of higher order in $P$. For the first order, the resulting closed coupled equation of motions are  
\begin{eqnarray}
\frac{\rm d}{{\rm d}t}{\bf G}_0^{(1)}&=&-4P{\bf G}_0^{(0)}+2 \gamma_{sp} {\bf BX}_{0}^{(1)} \nonumber   \\
&&+4 {\rm Re} \left( {\bf t}_{X,1}^{(1)}\right)=0, \label{eq:perturb1stE1} \\
\frac{\rm d}{{\rm d}t}{\bf BX}_0^{(1)}&=&-(\gamma_{sp}+\gamma_S){\bf BX}_0^{(1)} + P {\bf G}_0^{(0)} \nonumber \\
&&+ \gamma_S {\bf DX}_0^{(1)}-2 {\rm Re}\left( {\bf t}_{X,1}^{(1)}\right)=0, \label{eq:perturb1stE2} \\
\frac{\rm d}{{\rm d}t}{\bf DX}_0^{(1)}&=& - \gamma_S {\bf DX}_0^{(1)}+ \gamma_{S} {\bf BX}_{0}^{(1)}+P{\bf G}_0^{(0)} \nonumber \\
&=&0, \label{eq:perturb1stE3} \\
\frac{\rm d}{{\rm d}t}{\bf t}_{X,1}^{(1)}&=&-\left(\frac{\kappa+ \gamma_{sp}+ \gamma_{S}+2 \Gamma_{\rm ph}}{2} \right){\bf t}_{X,1}^{(1)} \nonumber  \\
&& -\frac{2 g^2}{ \gamma_{sp}+  \gamma_{S}+2 \Gamma_{\rm ph}}{\rm Re}\left( {\bf t}_{X,1}^{(1)} \right) \nonumber \\
&& +g^2 {\bf BX}_0^{(1)}-g^2 {\bf G}_{1}^{(1)}=0, \label{eq:perturb1stE4} \\
\frac{\rm d}{{\rm d}t}{\bf G}_1^{(1)}&=&-\kappa {\bf G}_1^{(1)} +4 {\rm Re}\left( {\bf t}_{X,1}^{(1)} \right)=0, \label{eq:perturb1stE5}
\end{eqnarray} 
with the first-order normalization condition ${\bf N}_0^{(1)}=0$ which reads
\begin{eqnarray}
{\bf G}_0^{(1)}=-2{\bf BX}_0^{(1)}-2{\bf DX}_0^{(1)}. \label{eq:nomalization1}
\end{eqnarray}
%%%%%%%%%%%%
Since $\frac{\rm d}{{\rm d}t} \left({\bf G}_0^{(1)}+2{\bf BX}_0^{(1)}+2{\bf DX}_0^{(1)} \right)=0$ irrespective of the stationary conditions (from Eq.~(\ref{eq:nomalization1})), there are only two independent equations among Eqs.~(\ref{eq:perturb1stE1})-(\ref{eq:perturb1stE3}).
The first-order equations are written in matrix form as
\begin{eqnarray}
 && \left(
    \begin{array}{cccc}
      2\gamma_{sp} & 0 & 4 & 0 \\
     \gamma_S & -\gamma_S & 0 & 0 \\
      g^2 & 0 & -C_1 & -g^2 \\
     0  & 0 & 4 & -\kappa 
    \end{array}
  \right)
  \left(
   \begin{array}{c}
    {\bf BX}_{0}^{(1)} \\
    {\bf DX}_{0}^{(1)} \\
    {\bf t}_{X,1}^{(1)} \\
    {\bf G}_{1}^{(1)}
   \end{array}
  \right) \nonumber \\
 && \qquad \qquad =
   \left(
   \begin{array}{c}
   4P  {\bf G}_{0}^{(0)} \\
    -P {\bf G}_{0}^{(0)} \\
    0 \\
    0
   \end{array}
  \right),  \label{eq:perturb1stMatrix}
\end{eqnarray}
where the coefficient $C_1$ given by  
\begin{eqnarray}
C_1=\frac{\kappa +\gamma_{sp}+\gamma_S +2\Gamma_{\rm ph}}{2}+\frac{2g^2}{\gamma_{sp}+ \gamma_S +2 \Gamma_{\rm ph}}.
\end{eqnarray}
Now seen in an explicit form, the correlation function to first-order in $P$ is given by the source term of the zeroth order, ${\bf G}_0^{(0)}(=1)$, and we find
\begin{eqnarray}
  \left(
   \begin{array}{c}
    {\bf BX}_{0}^{(1)} \\
    {\bf DX}_{0}^{(1)} \\
    {\bf t}_{X,1}^{(1)} \\
    {\bf G}_{1}^{(1)}
   \end{array}
  \right)
  =
    \left(
   \begin{array}{c}
   \frac{8 g^2 + 2 C_1 \kappa}{
C_1 \gamma_{sp} \kappa + 2 g^2 (2 \gamma_{sp} + \kappa)}
 \\
   \frac{C_1 (2 \gamma_S + \gamma_{sp}) \kappa + 
   2 g^2 (4 \gamma_S + 
      2 \gamma_{sp} + \kappa)}{ \gamma_S (C_1 \gamma_{sp} \kappa \
+ 2 g^2 (2 \gamma_{sp} + \kappa))} 
\\
  \frac{2 g^2  \kappa}{
C_1 \gamma_{sp} \kappa + 2 g^2 (2 \gamma_{sp} + \kappa)}
 \\
    \frac{8 g^2 }{C_1 \gamma_{sp}  \kappa + 2 g^2 (2 \gamma_{sp} + \kappa)}
   \end{array}
  \right) \times P. \nonumber \\ \label{eq:Perturbation1st}
\end{eqnarray}

From Eqs.~(\ref{eq:balance}) and (\ref{eq:Perturbation1st}), and the assumption that the $P$-dependence is in the linear regime, 
we obtain the average photon number

\begin{eqnarray}
{\bf N}_1^{(1)} = \frac{8 g^2 P}{
C_1 \gamma_{sp} \kappa + 2 g^2 (2 \gamma_{sp} + \kappa)} .
\end{eqnarray}

Being similar to the result for QD SP emitters without cavities in Eqs.~(\ref{eq:rhoXX})-(\ref{eq:rhoDX}), we found that only the dark state population ${\bf DX}_0^{(1)}$ (and also ${\bf G}_0^{(1)}$ from Eq.~(\ref{eq:nomalization1})) is divergent at $\gamma_S \to 0$.
Of course, this perturbation analysis applies only in the weak pumping limit, and the low-order perturbation treatment becomes unreliable if the contribution becomes of order unity. 
However, the ratio ${\bf DX}_0^{(1)}/{\bf BX}_0^{(1)}$ can be arbitrarily large for small $\gamma_S$ even in the linear regime. 
In a similar manner, a strong enhancement also occurs in $g^{(2)}(0)$ as seen below.

By repeating carefully a similar analysis for the second-order correlation functions ${\bf O}_k^{(2)}$ and assuming a steady state, we obtain the following closed set of equations: 
\begin{widetext}
\begin{eqnarray}
 &&  \left(
    \begin{array}{ccccccccccccc}
      0 & 0 & 0 & 0 & 0  &{\scriptscriptstyle  2\gamma_{sp}} & 0 & 0& 0 & 2 & 2 & 0 & 0 \\ 
      0 & 0 & 0 & {\scriptscriptstyle -\gamma_S }& 0  &{\scriptscriptstyle  \gamma_S} & 0 & 0 & 0 & 0 & 0 & 0 & 0  \\ 
      0 & 0 & 0 & 0 & 0  & 0 & {\scriptscriptstyle 2\gamma_{sp} }& 0 & 0 & 0 & 0 & 2 & 2  \\ 
      0 &{\scriptscriptstyle  -g^2} & 0 & 0 & {\scriptscriptstyle g^2}  & {\scriptscriptstyle g^2} & 0 &{\scriptscriptstyle -C_3 -C_4} & {\scriptscriptstyle -C_4}& {\scriptscriptstyle -C_1+C_2} & {\scriptscriptstyle -C_2} & {\scriptscriptstyle C_2+2C_3+C_4} & {\scriptscriptstyle C_2+C_4 } \\ 
      0 & {\scriptscriptstyle -g^2} & 0 & 0 & {\scriptscriptstyle g^2}  & {\scriptscriptstyle g^2} & 0 & {\scriptscriptstyle -C_4} & {\scriptscriptstyle -C_3^\ast -C_4} & {\scriptscriptstyle -C_2} & {\scriptscriptstyle -C_1+C_2} & {\scriptscriptstyle C_2+C_4 } & {\scriptscriptstyle C_2+2C_3^\ast+C_4 }  \\ 
      0 & 0 & 0 & 0 & {\scriptscriptstyle -g^2}  & 0 & {\scriptscriptstyle g^2} & {\scriptscriptstyle C_3 +C_4 } & {\scriptscriptstyle C_4} & 0 & 0 & {\scriptscriptstyle -2C_3-C_4-C_5 } &{\scriptscriptstyle  -C_4 }  \\
      0 & 0 & 0 & 0 & {\scriptscriptstyle -g^2}  & 0 & {\scriptscriptstyle g^2} & {\scriptscriptstyle C_4 } & {\scriptscriptstyle C_3^\ast +C_4 } & 0 & 0 & {\scriptscriptstyle -C_4 } &{\scriptscriptstyle  -2C_3^\ast-C_4-C_5^\ast }  \\ 
      0 &{\scriptscriptstyle - \kappa} & 0 & 0 & {\scriptscriptstyle 2 \gamma_{sp}} & 0  & 0 & 2 & 2 & 2 & 2 & 0 & 0  \\
      0 & 0 & {\scriptscriptstyle \gamma_S } & 0 & {\scriptscriptstyle -\kappa -\gamma_{sp}-\gamma_S}  & 0 & 0 & -1 & -1 & 0 & 0 & 1 & 1  \\
      0 & 0 & {\scriptscriptstyle -\kappa - \gamma_S } & 0 &  {\scriptscriptstyle  \gamma_S}  & 0 & 0 & 0 & 0 & 0 & 0 & 0 & 0  \\
     {\scriptscriptstyle -g^2} & 0 & 0 & 0 & {\scriptscriptstyle 2 g^2}  & 0 & 0 &  {\scriptscriptstyle -C_3 -2 C_4 -C_6} & {\scriptscriptstyle -2 C_4} & 0 & 0 & {\scriptscriptstyle 2 C_3 +2 C_4} & {\scriptscriptstyle 2 C_4}  \\
     {\scriptscriptstyle -g^2} & 0 & 0 & 0 & {\scriptscriptstyle 2 g^2}  & 0 & 0 & {\scriptscriptstyle -2 C_4 }& {\scriptscriptstyle -C_3^\ast -2 C_4 -C_6} & 0 & 0 &{\scriptscriptstyle  2 C_4} & {\scriptscriptstyle 2 C_3^\ast +2 C_4} \\
     {\scriptscriptstyle -2 \kappa } & 0 & 0 & 0 & 0  & 0 & 0 & 4 & 4 & 0 & 0 & 0 & 0  
    \end{array}
  \right) \nonumber \\ 
&& \qquad \qquad   \qquad  \qquad \qquad \qquad 
  \times 
  \left(
   \begin{array}{c}
   {\bf G}_{2}^{(2)} \\
   {\bf G}_{1}^{(2)} \\
   {\bf DX}_{1}^{(2)} \\
   {\bf DX}_{0}^{(2)} \\
   {\bf BX}_{1}^{(2)} \\
   {\bf BX}_{0}^{(2)} \\
   {\bf XX}_{0}^{(2)} \\
   {\bf t}_{X,2}^{(2)} \\
   {\bf t}_{X,2}^{(2) \ast} \\
   {\bf t}_{X,1}^{(2)} \\
   {\bf t}_{X,1}^{(2) \ast} \\
   {\bf t}_{XX,1}^{(2)} \\
   {\bf t}_{XX,1}^{(2) \ast}
   \end{array}
  \right)
  =
   \left(
   \begin{array}{c}
   4  {\bf G}_{0}^{(1)} \\
    {\bf DX}_{0}^{(1)} - {\bf G}_{0}^{(1)} \\
   2 {\bf BX}_{0}^{(1)} +2 {\bf DX}_{0}^{(1)} \\
    \left(\frac{5}{2}- \frac{2 C_2}{\gamma_{sp}+ \gamma_S +2 \Gamma_{\rm ph}} \right) {\bf t}_{X,1}^{(1)}  \\
     \left(\frac{5}{2}- \frac{2 C_2}{\gamma_{sp}+ \gamma_S +2 \Gamma_{\rm ph}} \right) {\bf t}_{X,1}^{(1)} \\
     0 \\
     0 \\
      4  {\bf G}_{1}^{(1)} \\
      -  {\bf G}_{1}^{(1)} \\
      -  {\bf G}_{1}^{(1)} \\
        0 \\
        0 \\
        0
   \end{array} 
  \right) \times P, \label{eq:2ndPerturbation}
\end{eqnarray}
\end{widetext}
where the decay coefficients are 
\begin{eqnarray}
C_2&=&\frac{g^2 }{\gamma_{sp} +\gamma_S +2\Gamma_{\rm ph}}, \\
C_3&=& \frac{2g^2 }{\kappa+\gamma_{sp}  +2\Gamma_{\rm ph}-i \chi} \\
C_4&=&\frac{g^2 }{\kappa+  \gamma_{sp} +\gamma_S +2\Gamma_{\rm ph}}, \\
C_5&=& \frac{\kappa+3\gamma_{sp}  +2\Gamma_{\rm ph} + \gamma_S }{2}-i \chi \\
C_6&=&\frac{3\kappa+\gamma_{sp}  +\gamma_S  +2\Gamma_{\rm ph} }{2}.
\end{eqnarray}
A matrix inversion of Eq.~(\ref{eq:2ndPerturbation}) yields the second-order perturbation series with the first-order ones as source terms. 
Since the first-order source terms contain ${\bf DX}_0^{(1)}$ and ${\bf G}_0^{(1)}$ $\propto 1/\gamma_S$, the second-order correlation functions relevant to ${\bf N}_{2}^{(2)}$ ($=4{\rm Re}( {\bf t}_{X,2}^{(2)})/\kappa$ from Eq.~(\ref{eq:balance})) also contain the divergent factor  $\propto 1/\gamma_S$. 
The second-order normalization condition ${\bf N}_0^{(2)}=0$ reads
\begin{eqnarray}
{\bf G}_0^{(2)}=-2{\bf BX}_0^{(2)}-2{\bf DX}_0^{(2)}-{\bf XX}_0^{(2)}. 
\end{eqnarray} 

\begin{figure}[tb]
\begin{center} 
\includegraphics[scale=0.82]{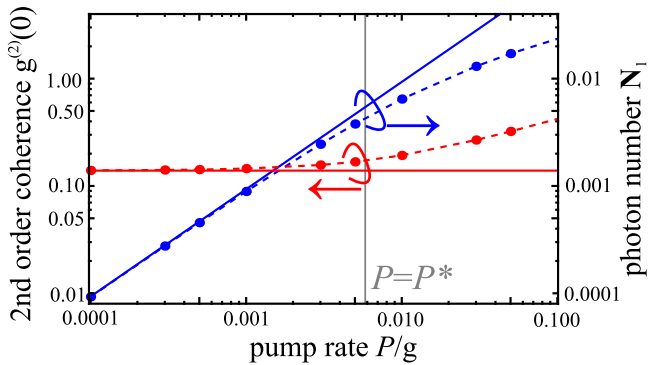} {}
\end{center} 
\vspace{0mm}
\caption{\label{fig:suppleB}  Second-order coherence $g^{(2)}(0)$ (blue) and mean cavity photon number ${\bf N}_{1}$ (red) are shown as a function of pumping rate $P/g$. The results evaluated by two different approaches, a perturbation analysis to their lowest order, $g^{(2)}(0) \approx {\bf N}_{2}^{(2)}/({\bf N}_{1}^{(1)})^2$ and ${\bf N}_{1}\approx {\bf N}_{1}^{(1)}$ (solid), and a direct numerical simulation of the QME (points), are compared. The dashed lines are guide for eye.   
We set $(g, \gamma_{sp}, \Gamma_{\rm ph}, \chi, \kappa, \gamma_{S})=(68, 10, 15, 2000, 200, 4.13)$ in $\mu$eV. The vertical grey line marks the definition of $P^*$}
\end{figure}

By repeating the same steps, it is in principle possible to obtain higher order correlation functions analytically in the liner weak-pump limit. 
However, third- and higher-order analysis requires much more laborious calculations, so we limit our calculation to the the second order. 
In order to check the validity of the perturbation result, in Fig.~\ref{fig:suppleB} we compare the analytic results for the mean photon number ${\bf N}_{1}={\bf N}_{1}^{(1)}$ and $g^{(2)}(0)={\bf N}_{2}^{(2)}/({\bf N}_{1}^{(1)})^2$ with the numerical QME results obtained after time evolutions until the system had reached the steady state. In the figure, we clearly see that the two methods give the same result, and the perturbation result is correct in the linear regime for weak pumping rate.

We now briefly note how we determined the crossover photon emission rate $I^\ast$ in the main text and in Fig.~\ref{fig2} and Fig.~\ref{fig3}. 
As seen from the QME result in Fig.~\ref{fig:suppleB}, $g^{(2)}(0)$ increases with $P$ while it is constant in the linear regime. 
This is natural since the increase in the cavity photon number directly leads to the increase in multiple photon emission events.
Therefore, real SP emitting devices should work in the linear regime to maintain a small value $g^{(2)}(0)$.
The crossover photon emission rate $I^\ast=\kappa \times {\bf N}_1(P=P^\ast)$ is the maximum cavity photon number for a pump rate at the crossover between the linear and non-linear regimes ($P=P^\ast$). This is defined more rigorously in our simulation as the condition where second order corrections amount to fifty percent of the first order corrections with respect to ${\bf N}_{1}$ and ${\bf BX}_0$. If ${\bf N}_{1}^{(2)}(P=P_1)=0.5 \times {\bf N}_{1}^{(1)}(P=P_1)$ and ${\bf BX}_{0}^{(2)}(P=P_2)=0.5 \times {\bf BX}_{0}^{(1)}(P=P_2)$, the smaller pump rate between $P_1$ and $P_2$, at which the nonlinear $P$-dependence  becomes apparent in $g^{(2)}(0)$, corresponds to $P^\ast$:
\begin{eqnarray}
P^\ast = 0.5 \times \min \left( \frac{{\bf N}_{1}^{(1)}/P}{{\bf N}_{1}^{(2)}/P^2}, \frac{{\bf BX}_{0}^{(1)}/P}{{\bf BX}_{0}^{(2)}/P^2} \right). \label{eq:Pmax}
\end{eqnarray}
The evaluation of this is possible within the second order perturbation analysis presented here, and we mark $P=P^\ast$ obtained by Eq.~(\ref{eq:Pmax}) in Fig.~\ref{fig:suppleB}.
We find that the calculated $P^\ast$ agreemees well with the numerical QME result, where the nonlinear $P-$dependence in ${\bf N}_1$ and a deviation from the weak pump limit of $g^{(2)}(0)$ are found. 
$I^\ast$ evaluated in this way therefore gives a good estimate for a maximum available SP emission rate, above which the QD emitters have a degraded single photon purity.

Finally, we give analytic expressions for the correlation functions for a system without DX states, DX1 and DX2, and the dark pumping path via the DX states. Comparison between the results obtained with and without the dark path can be used to evaluate the increase in $g^{(2)}(0)$ due to the path.
In the no dark path case, the first-order normalization condition in Eq.~(\ref{eq:nomalization1}) is replaced by
\begin{eqnarray}
{\bf G}_0^{(1)}=-2{\bf BX}_0^{(1)}, \label{eq:nomalization1without}
\end{eqnarray}
the first-order equation in Eq.~(\ref{eq:nomalization1}) by
\begin{eqnarray}
 && \left(
    \begin{array}{ccc}
      2\gamma_{sp} & 4 & 0 \\
      g^2  & -C_1 & -g^2 \\
      0 & 4 & -\kappa 
    \end{array}
  \right)
  \left(
   \begin{array}{c}
    {\bf BX}_{0}^{(1)} \\
    {\bf t}_{X,1}^{(1)} \\
    {\bf G}_{1}^{(1)}
   \end{array}
  \right) \nonumber \\
 && \qquad \qquad =
   \left(
   \begin{array}{c}
   2P  {\bf G}_{0}^{(0)} \\
    0 \\
    0
   \end{array}
  \right),  \label{eq:perturb1stMatrixwithout}
\end{eqnarray}
and the second-order equation in Eq.~(\ref{eq:2ndPerturbation}) by
\begin{widetext}
\begin{eqnarray}
 &&  \left(
    \begin{array}{ccccccccccc}
      0 & 0  & 0  &{\scriptstyle  2\gamma_{sp}} & 0 & 0& 0 & 2 & 2 & 0 & 0 \\ 
      0 & 0 & 0  & 0 & {\scriptstyle 2\gamma_{sp} }& 0 & 0 & 0 & 0 & 2 & 2  \\ 
      0 &{\scriptstyle  -g^2}  & {\scriptstyle g^2}  & {\scriptstyle g^2} & 0 &{\scriptstyle -C_3 -C_4} & {\scriptstyle -C_4}& {\scriptstyle -C_1+C_2} & {\scriptstyle -C_2} & {\scriptstyle C_2+2C_3+C_4} & {\scriptstyle C_2+C_4 } \\ 
      0 & {\scriptstyle -g^2}  & {\scriptstyle g^2}  & {\scriptstyle g^2} & 0 & {\scriptstyle -C_4} & {\scriptstyle -C_3^\ast -C_4} & {\scriptstyle -C_2} & {\scriptstyle -C_1+C_2} & {\scriptstyle C_2+C_4 } & {\scriptstyle C_2+2C_3^\ast+C_4 }  \\ 
      0 & 0  & {\scriptstyle -g^2}  & 0 & {\scriptstyle g^2} & {\scriptstyle C_3 +C_4 } & {\scriptstyle C_4} & 0 & 0 & {\scriptstyle -2C_3-C_4-C_5 } &{\scriptstyle  -C_4 }  \\
      0 & 0 & {\scriptstyle -g^2}  & 0 & {\scriptstyle g^2} & {\scriptstyle C_4 } & {\scriptstyle C_3^\ast +C_4 } & 0 & 0 & {\scriptstyle -C_4 } &{\scriptstyle  -2C_3^\ast-C_4-C_5^\ast }  \\ 
      0 &{\scriptstyle - \kappa}& {\scriptstyle 2 \gamma_{sp}} & 0  & 0 & 2 & 2 & 2 & 2 & 0 & 0  \\
      0 & 0 & {\scriptstyle -\kappa -\gamma_{sp}}  & 0 & 0 & -1 & -1 & 0 & 0 & 1 & 1  \\
     {\scriptstyle -g^2} & 0 & {\scriptstyle 2 g^2}  & 0 & 0 &  {\scriptstyle -C_3 -2 C_4 -C_6} & {\scriptstyle -2 C_4} & 0 & 0 & {\scriptstyle 2 C_3 +2 C_4} & {\scriptstyle 2 C_4}  \\
     {\scriptstyle -g^2} & 0 & {\scriptstyle 2 g^2}  & 0 & 0 & {\scriptstyle -2 C_4 }& {\scriptstyle -C_3^\ast -2 C_4 -C_6} & 0 & 0 &{\scriptstyle  2 C_4} & {\scriptstyle 2 C_3^\ast +2 C_4} \\
     {\scriptstyle -2 \kappa } & 0 & 0  & 0 & 0 & 4 & 4 & 0 & 0 & 0 & 0  
    \end{array}
  \right) \nonumber \\ 
&& \qquad \qquad   \qquad  \qquad \qquad \qquad 
  \times 
  \left(
   \begin{array}{c}
   {\bf G}_{2}^{(2)} \\
   {\bf G}_{1}^{(2)} \\
   {\bf BX}_{1}^{(2)} \\
   {\bf BX}_{0}^{(2)} \\
   {\bf XX}_{0}^{(2)} \\
   {\bf t}_{X,2}^{(2)} \\
   {\bf t}_{X,2}^{(2) \ast} \\
   {\bf t}_{X,1}^{(2)} \\
   {\bf t}_{X,1}^{(2) \ast} \\
   {\bf t}_{XX,1}^{(2)} \\
   {\bf t}_{XX,1}^{(2) \ast}
   \end{array}
  \right)
  =
   \left(
   \begin{array}{c}
   2  {\bf G}_{0}^{(1)} \\
   2 {\bf BX}_{0}^{(1)} \\
    \left(\frac{3}{2}- \frac{2 C_2}{\gamma_{sp} +2 \Gamma_{\rm ph}} \right) {\bf t}_{X,1}^{(1)}  \\
     \left(\frac{3}{2}- \frac{2 C_2}{\gamma_{sp}+2 \Gamma_{\rm ph}} \right) {\bf t}_{X,1}^{(1)} \\
     0 \\
     0 \\
     2  {\bf G}_{1}^{(1)} \\
      -  {\bf G}_{1}^{(1)} \\
        0 \\
        0 \\
        0
   \end{array} 
  \right) \times P, \label{eq:2ndPerturbationwithout}
\end{eqnarray}
\end{widetext}
where all the coefficients $C_l$ are evaluated at $\gamma_S=0$.
The crossover photon emission rate $I^\ast$ in this case is evaluated again by Eq.~(\ref{eq:Pmax}) in Fig.~\ref{fig2} and Fig.~\ref{fig3}.

\subsection{\label{suppleC} Rate equation analysis for SP emitters with cavities: the existence of an optimal cavity loss $\kappa$}
In this section, we shall see why there is an optimal cavity loss parameter minimizing $g^{(2)}(0)$ as found in Figs.~\ref{fig2} (b)-(e) in the main text. 

As referred to in the main text, for good cavities with small $\kappa$, photons accumulating in cavity increase the multi-photon probability and hence $g^{(2)}(0)$. Therefore, $g^{(2)}(0)$ decreases with increasing $\kappa$. 
This is similar to cavity QED of two-level systems (cavity loss $\kappa$, spontaneous emission $\gamma$, coupling $g$). An expression for $g^{(2)}(0)$ is given by 
\begin{eqnarray}
g^{(2)}_{\rm TLS}(0)=2\frac{\kappa+\gamma}{3 \kappa +\gamma}
\frac{4g^2+\kappa \gamma}{4g^2+\kappa (\kappa + \gamma)},
\end{eqnarray}
in \cite{Gartner}, for which one finds $g^{(2)}_{\rm TLS}(0)=2$ as $\kappa \to 0$.

\begin{widetext}
\begin{figure*}[tb]
\begin{center} 
\includegraphics[scale=0.50]{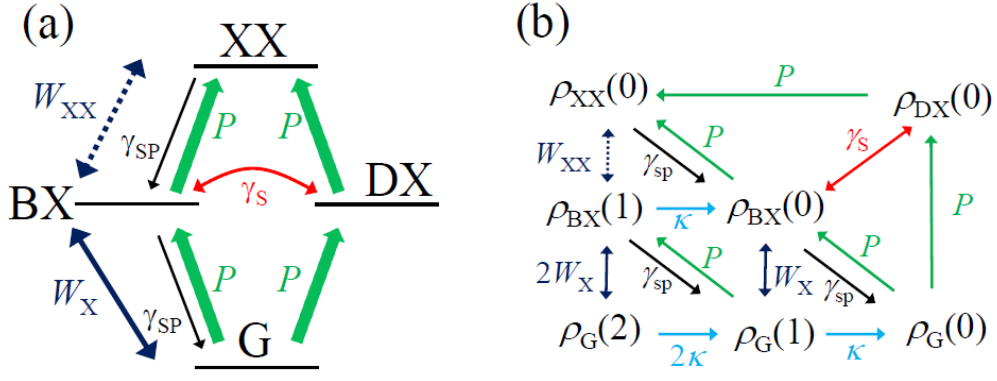} {}
\vspace{0mm}
\caption{\label{fig:suppleC} (a) Simple QD model: $W_{\rm XX}$ and $W_{\rm X}$ are the rates of the radiative transitions between XX state and BX state and between BX and G states into the cavity mode. The same notation is used for other parameters as in the main text. (b) Flow of the probabilities (diagonal parts of the density matrix) in the QD-cavity system. Assuming the weak pump limit (linear regime), elements of the density matrix relevant to the study is limited to a biexciton state with no photons, $\rho_{\rm XX}(0)$, bright exciton states with zero and one photons, $\rho_{\rm BX}(0)$ and $\rho_{\rm BX}(1)$, empty states with zero, one, two photons, $\rho_{\rm G}(0)$, $\rho_{\rm G}(1)$, and $\rho_{\rm G}(2)$, and a dark exciton state with no photons, $\rho_{\rm DX}(0)$.}
\end{center} 
\end{figure*}
\end{widetext}

Hereafter, we will focus on bad cavities in the weak coupling limit $\kappa \gg g$ to show the existence of the optimal $\kappa$. 
A rate equation analysis is presented by using the simplified QD model as shown in Fig.~\ref{fig:suppleC} (a) for a qualitative discussion.
The cavity mode is assumed to be in resonance with the exciton transition (BX $\leftrightarrow$ G). 
We denote the rates of the radiative transitions into the cavity mode by $W_{\rm X}$ for the exciton transition and $W_{\rm XX}$ for the biexciton transition.

\begin{widetext}
\begin{figure*}[tb]
\begin{center}
\includegraphics[scale=0.55]{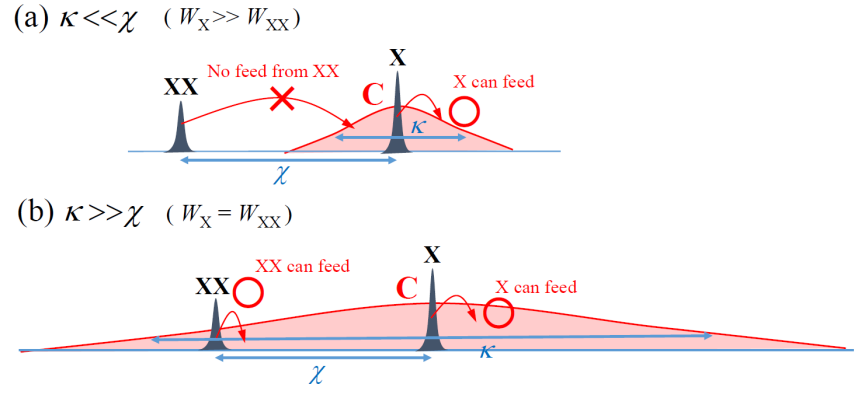} 
\vspace{-2mm}
\caption{\label{fig:suppleC2} Spectral profiles of the QD-bad cavity system for a schematic explanation of the transition rates $W_{\rm XX}$ and $W_{\rm X}$:  (a) $W_{\rm X} \gg W_{\rm XX}$ for $\kappa \ll \chi$, and (b) $W_{\rm X} = W_{\rm XX}$ for $\kappa \gg \chi$.} 
\end{center} 
\end{figure*}
\end{widetext}

Since we are focusing on the weak pump regime, we take into account only the density matrix elements limited to those for a biexciton state with no photons, $\rho_{\rm XX}(0)$, bright exciton states with zero and one photons, $\rho_{\rm BX}(0)$ and $\rho_{\rm BX}(1)$, empty states with zero, one, two photons, $\rho_{\rm G}(0)$, $\rho_{\rm G}(1)$, and $\rho_{\rm G}(2)$, and a dark exciton state with no photons $\rho_{\rm DX}(0)$.
From the probability flows in Fig.~\ref{fig:suppleC} (b), 
we obtain the following set of rate equations: 
\begin{widetext}
\begin{eqnarray}
\dot{\rho}_{XX}(0)&=& -(W_{XX}+\gamma_{sp})\rho_{XX}(0)  +P \rho_{BX}(0)+P \rho_{ DX}(0)+W_{XX}\rho_{BX}(1),\\
\dot{\rho}_{BX}(1)&=& -(2W_{X}+W_{XX}+\gamma_{sp}+\kappa)\rho_{BX}(1) +W_{XX} \rho_{XX}(0)+P \rho_{G}(1)+2W_{X}\rho_{G}(2),\\
\dot{\rho}_{BX}(0)&=& -(W_{X}+\gamma_{sp}+\gamma_S+P)\rho_{BX}(0)+\kappa \rho_{BX}(1) +\gamma_{sp} \rho_{XX}(0)+P \rho_{G}(0) +\gamma_S \rho_{ DX}(0) +W_{X} \rho_{G}(1), \\
\dot{\rho}_{G}(2)&=& -(2\kappa+2W_{X}) \rho_{G}(2)+2W_{X} \rho_{BX}(1),\\
\dot{\rho}_{G}(1)&=&-(\kappa +P+W_X) \rho_{G}(1)+W_{X} \rho_{BX}(0) +\gamma_{sp} \rho_{BX}(1)+2 \kappa \rho_{G}(2), \\
\dot{\rho}_{G}(0)&=&-2P \rho_{G}(0) +\kappa \rho_{G}(1) +\gamma_{sp} \rho_{BX} (0), \\
\dot{\rho}_{ DX}(0)&=& -(P+\gamma_S)  \rho_{ DX}(0) +P\rho_G (0) +\gamma_S \rho_{BX}(0).
\end{eqnarray}
\end{widetext}
The rate equations is solved with the normalization condition, $\rho_{\rm XX}(0)+\rho_{\rm BX}(0)+\rho_{\rm BX}(1)+\rho_{\rm G}(0)+\rho_{\rm G}(1)+\rho_{\rm G}(2)+\rho_{\rm DX}(0)=1$.
For large $\kappa$ where $\kappa \gg (\gamma_{sp}, \gamma_S) \gg(W_X,W_{XX})$, we have for the steady state, 
\begin{eqnarray}
g^{(2)}(0)= 2 \frac{W_{XX}}{W_{X}}\left( 1+\frac{\gamma_{sp}}{4 \gamma_S}\right)
\end{eqnarray}
to the lowest order in $1/\kappa$. Here we have taken the limit $\kappa \to \infty$ after taking $P \to 0$. 
We find again the same enhancement factor of $1+\frac{\gamma_{sp}}{4 \gamma_S}$ in the expression (see Eq.~(\ref{eq:g2simple}) in the main text.)
The prefactor is determined by the ratio of the transition rates $2W_{\rm XX}/W_{\rm X}$ which depends on $\kappa$ as we shall see below.
If $\kappa$ is larger than $\gamma_{sp}$, $\gamma_S$, and $g$, but $\kappa \ll \chi$ as shown in Fig.~\ref{fig:suppleC2}(a), 
the transition probabilities are given by $W_{\rm X} =2g^2/\kappa$ and $W_{\rm X} =2g^2 \kappa /(\kappa^2 + \chi^2)$. In this case, the prefactor 
\begin{eqnarray}
2W_{\rm XX}/W_{\rm X} =2 \frac{\kappa^2}{\chi^2 +\kappa^2}
\end{eqnarray}
is small. For large $\chi \gg \kappa$, $g^{(2)}(0)$ vanishes, being 
analogous to the result in Eq.~(\ref{eq:g2simple}) in the main text.
On the other hand, if $\kappa$ is larger than $\gamma_{sp}$, $\gamma_S$, and $g$ and $\kappa \gg \chi$ as shown in Fig.~\ref{fig:suppleC2}(b), we find $W_{\rm X} =W_{\rm XX} =2g^2/\kappa$ and the prefactor equals two.
According to this result, for bad cavities $\kappa \gg (g,\gamma_S,\gamma_{sp})$, $g^{(2)}(0)$ increases with $\kappa$ until $\kappa \sim \chi$, where it begins to saturate at a value of $2 (1+\frac{\gamma_{sp}}{4 \gamma_S})$.

Now it is clear that there should be a minimum in $g^{(2)}(0)$ as a function of $\kappa$, from the 2 results: (i) $g^{(2)}(0)$ decreases with increasing $\kappa$ for small $\kappa$, and (ii) increases with $\kappa$ for large $\kappa$.
The optimal value of $\kappa \equiv \kappa_{\rm opt, 1}$ is found in a range $g<\kappa_{\rm opt, 1}<\chi$. 

\begin{thebibliography}{99}
\bibitem{note1} Following a discussion in Sec. 2.3 of \cite{Carmichael}, the expression is obtained by $g^{(2)}(0)= \langle \colon \hat{I} \hat{I} \colon \rangle/\langle  \hat{I} \rangle ^2$ where $\hat{I} \equiv \gamma_X \hat{S}_{X}^{+}\hat{S}_{X}^{-}+f_{XX} \gamma_{XX} \hat{S}_{XX}^{+}\hat{S}_{XX}^{-}$ is the operator for the total emission rate into the energy window, $\hat{S}_{X}^{+}=(\hat{S}_{X}^{-})^\dagger \equiv |BX \rangle \langle G | $, and $\hat{S}_{XX}^{+}=(\hat{S}_{XX}^{-})^\dagger \equiv |XX \rangle \langle BX | $. In the normal ordered product, $\langle \colon \hat{I} \hat{I} \colon \rangle = f_{XX} \gamma_{XX}\gamma_X \langle \hat{S}_{XX}^{+}\hat{S}_{X}^{+} \hat{S}_{X}^{-} \hat{S}_{XX}^{-}  \rangle$ is the only term remained. In the denominator, the dominant linear-order term in $P$ is $\langle  \hat{I} \rangle = \gamma_X  \langle  \hat{S}_{X}^{+}\hat{S}_{X}^{-} \rangle $.
\bibitem{note2} In an ideal case $\gamma_S=0$, we have $\rho_{XX}/\rho_{BX}^2 \approx \gamma_X^2/(2 \gamma_{XX} P)$ indicating $P=0$ corresponds to a singular point of $g^{2}(0)$.
\bibitem{note4}  Since an identification of the exciton state for the QD emission has not been done, there is a possibility that the photon emission was due to charged excitons in Ref.~\cite{Nomura}. A dissociation of the values of $g^{(2)}(0)$ between our simulation and the experiments~\cite{Nomura} suggests the possibility.
\end{thebibliography}
\end{document}